\documentclass[prb,twocolumn,showpacs,preprintnumbers,amsmath,amssymb,prb]{revtex4}

\usepackage{graphicx}
\usepackage{dcolumn}
\usepackage{bm}

\begin{document}


\title{Evolution of the Surface Structures on SrTiO$_3$(110) Tuned by Ti or Sr Concentration}

\author{Zhiming Wang, Fang Yang, Zhiqiang Zhang, Yuanyuan Tang, Jiagui Feng, Kehui Wu, Qinlin Guo, and Jiandong Guo}
 \email{jdguo@aphy.iphy.ac.cn}
 \affiliation{Beijing National Laboratory for Condensed-Matter Physics \& Institute of Physics, Chinese Academy of Sciences, Beijing 100190, P. R. China}
\date{\today}

\begin{abstract}

The surface structure of the SrTiO$_3$(110) polar surface is studied by scanning tunneling microscopy and X-ray photoelectron spectroscopy. Monophased reconstructions in (5$\times$1), (4$\times$1), (2$\times$8), and (6$\times$8) are obtained, respectively, and the evolution between these phases can be tuned reversibly by adjusting the Ar$^{+}$ sputtering dose or the amount of Sr/Ti evaporation. Upon annealing, the surface reaches the thermodynamic equilibrium that is determined by the surface metal concentration. The different electronic structures and absorption behaviors of the surface with different reconstructions are investigated. 
\end{abstract}

\pacs{68.47.Gh, 68.37.Ef, 68.35.Md, 82.80.Pv, 68.35.Rh}
\keywords{oxide surface, scanning tunneling microscopy, photoelectron spectroscopy, structural phase transition, surface segregation}
\maketitle

\section{\label{sec:level1}Introduction}

As a typical family of strongly-correlated electron materials, transition metal oxides (TMOs) possess immense novel functionalities that have drawn intensive interest in the fields of fundamental condensed matter physics, materials science, and electronic applications. The signature of TMOs is the collective electron behaviors determined by the delicate equilibrium of multiple low energy excitation states or competing ground states that are sensitively influenced by doping, strain induction, or the application of external field. In the TMOs with reduced dimensionality, broken symmetry, or spatial confinement, the coupling between different degrees of freedom is normally enhanced and exotic phenomena emerge. A tremendous amount of evidence has shown that thin films, superlattices and heterostructures of TMOs display an even richer diversity of remarkable properties that are related, but not identical to the bulk phenomena\cite{Heber}. One of the important and intriguing discoveries is the quasi-two-dimensional electron gas (2DEG) formed at the interface between SrTiO$_3$ and LaAlO$_3$ \cite{Nature04 Hwang}, which might be the basis of the transformative concept for the new generation of electronic devices. Additionally, Hwang \textit{et al.} found that the formation of 2DEG critically depends on the atomic arrangement at the interface -- it can be formed only if the SrTiO$_3$ substrate is terminated by the TiO$_2$ layer, while the interface is insulating when the substrate exposes the SrO layer \cite{Nature04 Hwang}. Therefore the fabrication of artificial oxide heterostructures with desired functionalities must be controllable with atomic precision. Among all the techniques for the growth of low-dimensional materials, pulsed laser deposition (PLD) and oxide molecular beam epitaxy (OMBE), developed with the intensive efforts to grow the high-temperature superconductor films over the past two decades, have been demonstrated to be effective for the tailored construction of multilayered oxide films \cite{H. M. Christen,I. Bozovic growth, Eckstein, D. G. Schlom, Nature04 Hwang, C. H. Ahn, K. Miyano}. Besides the capability to achieve the precision of monoatomic layer along the growth direction, it is also promising in tuning the in-plane microscopic structures in atomic scale. 

Single crystalline SrTiO$_3$ is widely used as the epitaxial growth substrate for a variety of complex oxide films and heterostructures. By selective chemical etching of the SrO layer followed by thermal annealing, atomically well-defined SrTiO$_3$(001) surface can be achieved \cite{Science94, APL98 Koster}. Microscopic studies \cite{SS542-177} showed a series of surface reconstructions c(2$\times$2), (2$\times$2), c(4$\times$2), two kinds of c(4$\times$4), (4$\times$4), ($\sqrt{5}\times\sqrt{5}$)-R26.6$^{o}$, and ($\sqrt{13}\times\sqrt{13}$)-R33.7$^{o}$, depending on the annealing temperatures in vacuum. The Sr adatom model consisting of ordered Sr adatoms on a TiO$_{2}$-terminated layer has been proposed based on the first-principles total-energy calculation \cite{PRL86-1801}. A transmission electron microscopy (TEM) study showed different surface structures that appear to be exceptionally stable in air \cite{SS526-107}. The stability of the surface phases strongly correlates to the stoichiometry \cite{SS425-343, SS516-33, SS600-L129, PRB75-205429}. Different reconstructions or even one-dimensional nanolines can be formed with varied surface chemical concentration ratio of Sr to Ti or the degree of oxygen deficiency. It has also been revealed that the TiO$_{2}$ termination is stable and predominantly exposes on the chemically etched SrTiO$_3$(001) surface, while the epitaxially grown SrO layer is unstable and cannot cover the surface fully \cite{APL91-101910}.

Along (110) direction, the SrTiO$_3$ single crystal is composed of alternately stacked (SrTiO)$^{4+}$ and (O$_2$)$^{4-}$ atomic layers, which creates a macroscopic dipole perpendicular to the surface \cite{JPCM00 Noguera, RPP08 Noguera}. Therefore the surface is inherently unstable. On one hand, this means that manipulating the polarity provides us an additional degree of freedom to tune the properties of the oxide multilayers \cite{LaVO3}. \textquotedblleft Electronic reconstruction\textquotedblright at the polar and non-polar oxide interface has been proposed recently \cite{electronic reconstruction}. On the other hand, in order to cancel the surface polarity and thus making the electrostatic energy converged, a polar surface bears high reconstruction instability \cite{RPP08 Noguera}. Besides several (1$\times$1) terminations proposed by the \textit{ab initio} calculations \cite{PRB03 Noguera}, a large family of reconstruction has been observed on SrTiO$_3$(110) surface. The surface oxygen vacancies can induce new long-range orders that differ from that in the bulk \cite{APL05 Hwang, PRB69-035408, SS566-231}. Brunen and Zegenhagen \cite{SS97 Brunen} observed (2$\times$5), (3$\times$4), (4$\times$4), (4$\times$7), and (6$\times$4) reconstructions by scanning tunneling microscopy (STM), low energy electron diffraction (LEED), and Auger electron spectroscopy (AES) measurements and found that the surface Sr concentration increases with the elevated annealing temperature. Bando \textsl{et al.} \cite{JVTB95 Bando} obtained (5$\times$2) and c(2$\times$6) reconstructed surfaces exhibiting metallic characteristics. Recently, M. R. Castell \textsl{et al.} \cite{PRB08 Castell} obtained an (n$\times$1) (3$\leq$n$\leq$6) family of reconstruction at different annealing temperatures. Their AES analyses revealed that the (3$\times$1) and (4$\times$1) were Ti-enriched, while the (6$\times$1) was Sr-enriched. Combined with the transmission electron diffraction and density functional theory studies, they proposed a corner-sharing TiO$_4$ tetrahedra model \cite{nm10 L.D. Marks}.

Our previous studies showed that the thermodynamically stable SrTiO and O-terminations on the SrTiO$_3$(110) surface can be obtained by Ar$^+$ sputtering followed by annealing in ultra high vacuum (UHV) \cite{Zhiming Wang APL09}. In this paper we establish the phase diagram of the SrTiO$_3$(110) surface in a wide scale, covering from the (5$\times$1)- to the (4$\times$1)-reconstructed surface of SrTiO termination, then to the (2$\times$8)- and finally to the (6$\times$8)-reconstructed surface of O termination. The concentration ratio of surface metal cations, [Ti]/[Sr], increases when the reconstruction evolves between the above phases sequentially as determined by the X-ray photoelectron spectroscopy (XPS) analyses. These reconstruction phases can be selected by adjusting the sputtering dose or the amount of Sr/Ti adsorption onto the surface. It is revealed that the thermodynamic stability of the reconstruction phases is determined by [Ti]/[Sr] on the surface, which is influenced by the diffusion of under-coordinated Ti induced by sputtering towards the surface upon annealing the as-sputtered sample or directly varied by evaporating Sr and Ti metals onto the surface. The properties of the surface with different reconstructions are also examined.

This paper is organized as follows: after the description of sample preparation and characterization in Sec. II, Sec. III reports how the surface reconstruction and composition are tuned by Ar$^+$ sputtering followed by UHV annealing, as well as by adsorption of Sr/Ti metals (also followed by UHV annealing) that is more direct and easier to control. We are then enabled to establish the phase diagram of the surface in a wide range. In Sec. IV, we discuss that the Ti diffuses towards the surface upon annealing and is responsible for the selective stabilization of different phases. Furthermore, we study the influence of different surface reconstructions on the electronic structures and metal adsorption behaviors. Finally the summary is given in Sec. V.

\section{\label{sec:level2}Experimental}

The experiments were performed in two separate UHV combined systems. The reconstruction phases and the surface chemical compositions of the sample were characterized in an Omicron system equipped with variable temperature STM, LEED and XPS. In XPS measurements, Mg \textit{K}$\alpha$ radiation and a pass energy of 20 eV were used. The binding energy was calibrated by using the copper metal (Cu~2\textsl{p}$_{3/2}$, energy of 932.9 eV relative to the Fermi level). In the other system equipped with low-temperature STM and reflective high-energy electron diffraction (RHEED), we carried out detailed and systematic imaging with high resolution in the real space. The base pressure of both systems was better than 1$\times$10$^{-10}$ mbar.

\begin {figure}[t]
 \includegraphics [width=2.8in,clip] {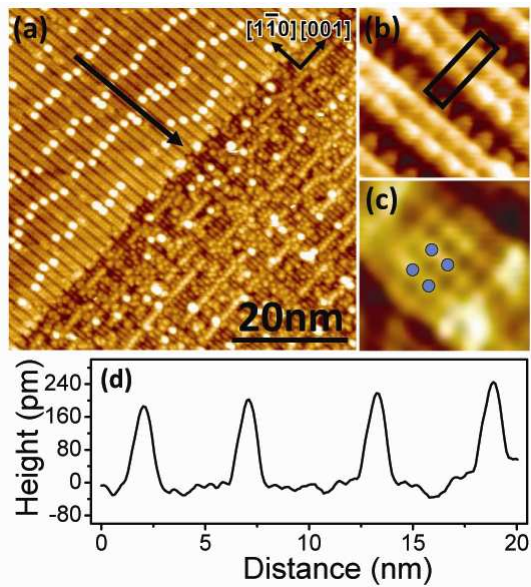}
 \caption{
(Color online) (a) An STM image (1.5~V/20~pA) of the SrTiO$_3$(110) surface with SrTiO and O terminations coexisting with each other. (b) and (c) High resolution STM images of the SrTiO (4$\times$4~nm$^2$, 1.0~V/200~pA) and O (2.6$\times$2.6~nm$^2$, 1.2~V/20~pA) terminations, respectively. The unit cell of the (4$\times$1) reconstruction is labeled in the inset. (d) The line profile along the arrow in (a), indicating an $\sim$\textquotedblleft $\times$10\textquotedblright periodicity of the clusters along [1$\overline{1}$0].
}
\end{figure}

Nd-doped (0.7~wt$\%$) SrTiO$_3$(110) single crystal (12~mm$\times$3~mm) were purchased from Hefei KMT Company, China. After loaded into the system, the as-received sample was sputtered with Ar$^+$ beam at room temperature (RT) followed by annealing in UHV. The energy of the ion beam was varied from 0.5 to 2~keV without any difference detected. The sample was heated by resistively passing a direct current through the crystal and the temperature was measured with an optical pyrometer. The annealing temperature was 1000 $^{o}$C for 1 hour unless otherwise specified. The maximum pressure during annealing did not exceed 2$\times$10$^{-9}$ mbar. Ti metal was evaporated by using an electron beam evaporator and Sr metal was evaporated with a low temperature effusion cell. The flux of Sr and Ti was calibrated by monitoring the RHEED intensity oscillation during the homoepitaxial growth of SrTiO$_3$(110) thin films (1~ML=4.64$\times$10$^{14}$~atoms/cm$^2$).

\section{\label{sec:level2}Results}
\subsection{Ar$^+$ Sputtering}

\begin {figure}[t]
 \includegraphics [width=2.3in,clip] {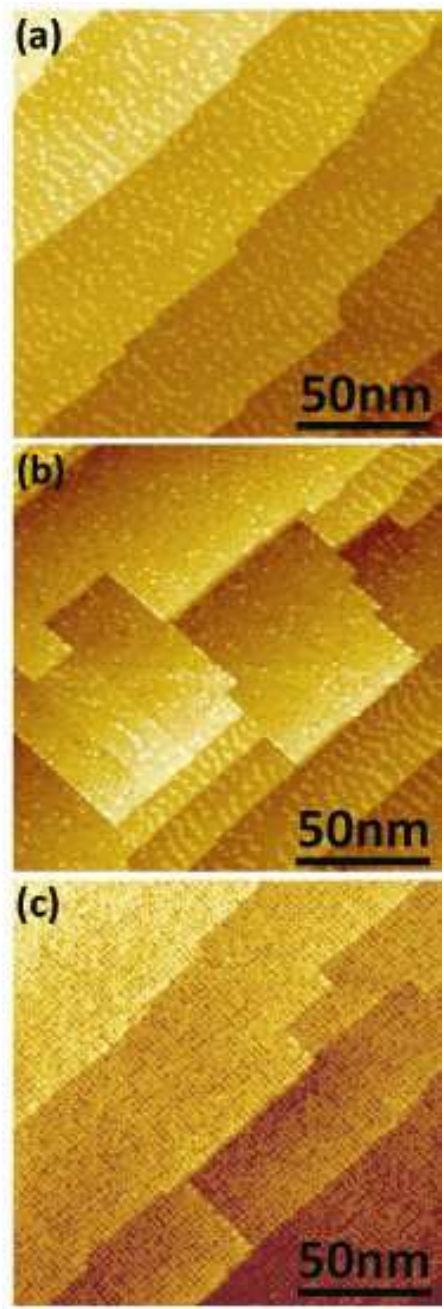}
 \caption{
(Color online) STM images (3.5~V/20~pA) of the SrTiO$_3$(110) treated with different Ar$^+$ sputtering dose followed by annealing. As the sputtering dose increases, the surface evolves from (a) the single type of SrTiO termination to (b) the mixed SrTiO and O terminations, and to (c) the single type of O termination.
}
\end{figure}

Figure 1 (a) shows two kinds of domains coexisting with each other on the surface of SrTiO$_3$(110) single crystal after Ar$^+$ sputtering and UHV annealing. As shown in Fig.~1~(b), one type of the domains consists of periodic stripes along the [1$\overline{1}$0] direction, separated with a dark trench, and each contains two obvious bright rows of periodic dots. The long-range ordering is characterized with the lattice constant of $\sim$0.6~nm along [1$\overline{1}$0] and 1.6~nm along [001], respectively. Such a structure corresponds to the (4$\times$1)-reconstructed SrTiO termination, which has been observed by another group \cite{PRB08 Castell, nm10 L.D. Marks}. There are clusters with uniform size adsorbed on the (4$\times$1) domain. The distribution of the clusters also exhibits a quasi-long-range ordering -- they form meandering lines along [001] direction and are separated by approximate \textquotedblleft $\times$10\textquotedblright periodicity along [1$\overline{1}$0] direction. The nature and the ordering mechanism of these \textquotedblleft magic\textquotedblright clusters are under investigations.

\begin {figure} [b]
 \includegraphics [width=2.8in,clip] {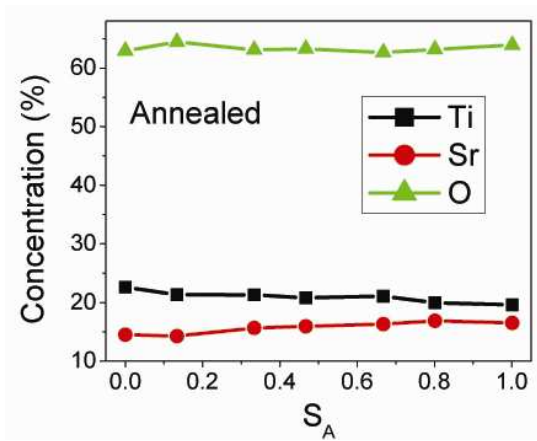}
 \caption{
(Color online) Surface chemical concentration of the SrTiO$_3$(110) surface after sputtering and annealing as determined by XPS. 
}
\end{figure}

As reported previously \cite{Zhiming Wang APL09}, the height between the (4$\times$1) and the other domain is about 0.138~nm, equivalent to the interspacing of adjacent SrTiO and O atomic layers along the [110] direction. The high-resolution STM image shows periodic protrusions along [1$\overline{1}$0] direction with the interspacing of $\sim$~0.3 nm, as marked by the dots in Fig.~1~(c). This is quite close to the value of $\sqrt{2}/2$a [a=0.3905 nm, the lattice constant of SrTiO$_3$(110) bulk crystal] that only possibly exists on the O-terminated (110) surface, consistently indicating that the domain corresponds to the O termination. The detailed structure of the O-terminated surface will be discussed in the following.

We are able to tune the termination type on the SrTiO$_3$(110) surface by adjusting the sputtering dose. As the sputtering dose increasing, the areal ratio of the SrTiO termination to the entire surface (S$_A$) decreases. The surface with a single type of termination (either SrTiO or O) can be obtained, as shown in Fig.~2~(a) and (c), respectively. We determine the surface atomic concentration with the analyses of XPS spectra following:

\begin{equation}
[M]= \frac{I_{M}/\Sigma_{M}}{I_{Sr}/\Sigma_{Sr}+I_{Ti}/\Sigma_{Ti}+I_{O}/\Sigma_{O}},
\end{equation}
where M donates Sr, Ti or O, I is the integrated intensity of the characteristic peak in XPS for each element, and $\Sigma$ is the element sensitivity factor that has been calibrated with vacuum fractured SrTiO$_3$(001) surface \cite{sensitivity factors}. As shown in Fig.~3, the surface chemical composition changes with the S$_A$ on the surface. The nominal composition of the SrTiO-terminated surface (S$_A$=1) is SrTi$_{1.19}$O$_{3.76}$, while it is SrTi$_{1.5}$O$_{4.5}$ for the O-terminated surface (S$_A$=0). The [Ti]/[Sr] ratio increases from 1.19 to 1.5 monotonically as the surface evolves from SrTiO to O termination. It also should be noted that the (110) surface treated with Ar$^+$ sputtering followed by UHV annealing is not oxygen-deficient for either termination, at least not more deficient than the vacuum fractured SrTiO$_3$(001) surface that has been used for calibration. 

\subsection{Adsorption of Sr and Ti metals}

Treated with Ar$^+$ sputtering followed by annealing, the SrTiO$_3$(110) surface is stable against high temperature or oxygen partial pressure \cite {Zhiming Wang APL09}. This indicates that the thermodynamic equilibrium has been established on the surface and the stabilization process of the termination layers is strongly related to the surface chemical composition. Therefore by directly evaporating Sr and/or Ti metals onto the surface, we can change the relative ratio of the termination types. Figure~4 shows the STM observations. On the SrTiO$_3$(110) surface with SrTiO and O terminations coexisting, a submonolayer of Sr metal is evaporated. After annealing we obtain the surface with the single type of SrTiO termination. To further evaporate 0.08~ML Ti metal followed by annealing, O-terminated islands are formed. As the dosage of Ti increasing, the area of O-terminated domains increases until the surface is fully covered by the single type of O termination. 

To evaporate Sr or Ti metal followed by annealing induces the formation of SrTiO or O termination on the (110) surface, respectively. It is a reversible process in which the S$_{A}$ value is determined by the dosage of Sr and Ti only, \textit{i.e.}, the surface [Ti]/[Sr] as detected by the XPS. Moreover, adjusting the evaporation dosage of Sr or Ti metal is practically easier and more precisely, which enables the detailed investigations of the evolution of different surface reconstructions beyond the simple identification of termination.

\begin {figure} [t]
 \includegraphics [width=2.2in,clip] {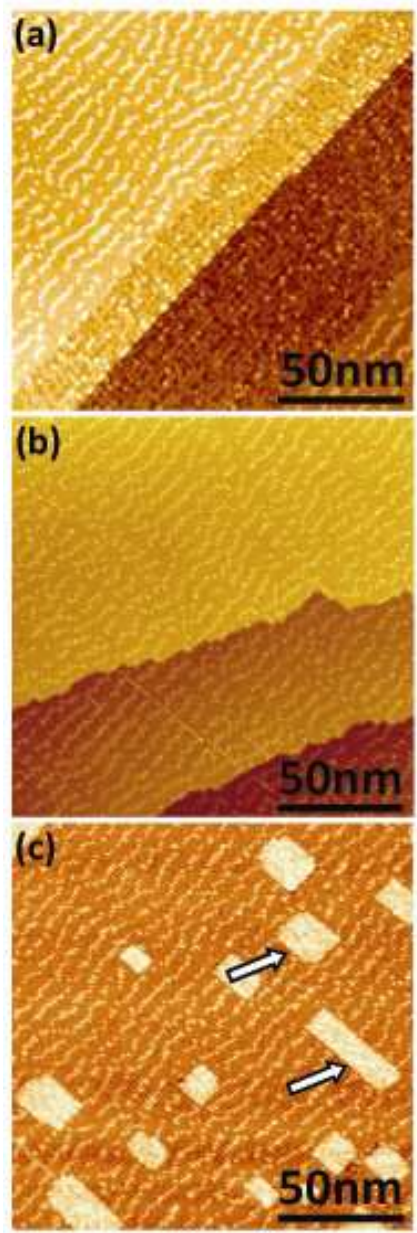}
 \caption{
(Color online) (a) An STM image (2.0 V/20 pA) of the (4$\times$1)-reconstructed SrTiO termination coexisting with the O termination. (b) The single type of SrTiO termination obtained by evaporating $\sim$~0.35~ML Sr onto the surface in (a) followed by annealing (2.0~V/20~pA). (c) The surface after depositing $\sim$0.08~ML Ti onto (b) followed by annealing (2.0~V/20~pA). The arrows indicate the O-terminated islands.
}
\end{figure}

On the SrTiO termination, we further increase the Sr concentration by evaporating a small amount of Sr metal onto the (4$\times$1) monophased surface. New stripes are observed after annealing, as shown in Fig.~5~(a). Each new stripe contains three bright rows of periodic dots along the [1$\overline{1}$0] direction. There also exist hole-like defects on the center row of the stripe, forming the quasi-ordering of \textquotedblleft $\times$10\textquotedblright along [1$\overline{1}$0]. The domain area of the new stripes enlarges with the Sr evaporation dosage increasing until a monophased surface formed with $\sim$0.15~ML Sr, as shown in Fig.~5 (b). The high-resolution STM image shows the new long-range ordering with the lattice constant of $\sim$0.6~nm along [1$\overline{1}$0] and $\sim$1.95~nm along [001], respectively, consistent with the RHEED patterns that clearly indicate the (5$\times$1) reconstruction. There are clusters adsorbed on the (5$\times$1) surface. Comparing to the clusters on the (4$\times$1) surface, their size and distribution are random and do not relate to the annealing temperature or the cleanliness of the chamber. 

Evaporating Ti metal onto the surface enlarges the (4$\times$1) domain over the (5$\times$1). The reversed phase transition from (5$\times$1) to (4$\times$1) reconstruction can be realized by evaporating $\sim$0.15~ML Ti followed by annealing. 

\begin {figure}[t]
 \includegraphics [width=3.4in,clip] {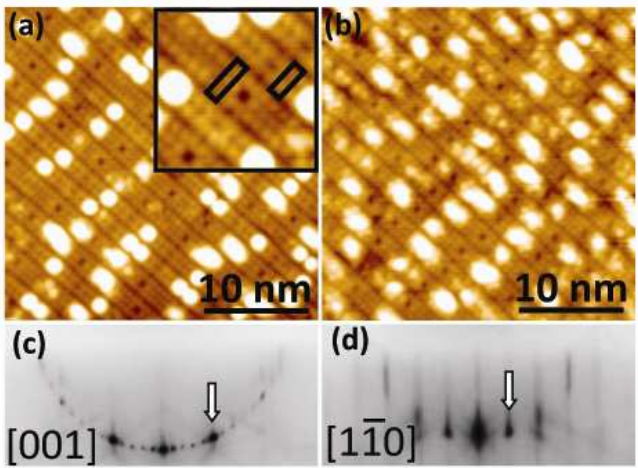}
 \caption{
(Color online) (a) The surface with the (5$\times$1) and (4$\times$1) phases coexisting with each other (1.6~V/20~pA). The unit cells are labeled in the inset with the rectangles, respectively. (b) The monophased (5$\times$1) surface (1.5~V/20~pA). (c) and (d) The RHEED patterns of the (5$\times$1) along [001] and [1$\overline{1}$0] directions, respectively. The integral diffractions are indicated with the arrows. 
}
\end{figure} 

As discussed above, Ti induces the formation of termination on the surface. By evaporating $\sim$0.75~ML Ti onto the (4$\times$1) monophased surface, we tune it to the single type of O termination after annealing. The high-resolution STM image is presented in Fig.~6~(a). Bright rows along [1$\overline{1}$0] are observed. They slightly pair up along [001], resulting in the narrow dark trench between every \textquotedblleft 2$\times$\textquotedblright stripe. Along the [1$\overline{1}$0] direction, each pair of the rows are separated by the dark kinks that are periodically distributed. Such that the STM measurements indicate the lattice constant of $\sim$0.75~nm along [001] and $\sim$2.3~nm along [1$\overline{1}$0], respectively. Such a long-range order corresponds to the (2$\times$8) reconstruction in relative to the lattice of O atomic layer, which is clearly visible in the RHEED patterns shown in Fig.~6~(b) and (c). The surface is densely adsorbed by clusters that can be divided into two categories by size. The small clusters are on the top of the paired rows with random distribution. The big clusters are over the trenches between the paired rows, and align to straight lines along [001]. These clusters seem to be intrinsic for the (2$\times$8) phase and are responsible for the charge compensation on the polar O-terminated SrTiO$_3$(110) surface, although the detailed mechanism is not understood yet.

\begin {figure}[t]
 \includegraphics [width=2.8in,clip] {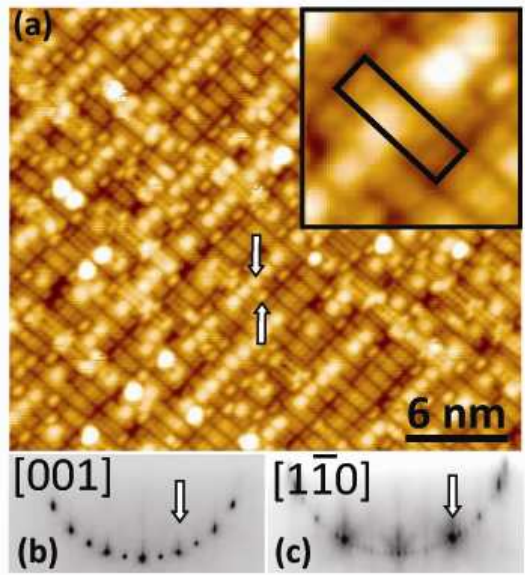}
 \caption{
(Color online) A high-resolution STM image (1.5~V/40~pA) of the monophased (2$\times$8) surface. Two kinds of clusters are labeled with up and down arrows, respectively. The (2$\times$8) unit cell is labeled on the zoom-in image in the inset. (b) and (c) The (2$\times$8) RHEED patterns along [001] and [1$\overline{1}$0] directions, respectively, with the integral diffractions indicated with the arrows.
}
\end{figure}

Further increasing the Ti dosage on the (2$\times$8) surface, a structural phase transition to (6$\times$8) occurs with the dosage of $\sim$0.1~ML followed by annealing. Comparing to the uniform row pairs in the (2$\times$8) phase, wide stripes with four protrusions in across appear and are arranged alternatively with the row pairs, as shown in Fig.~7~(a). The (6$\times$8) long-range ordering is clearly shown by the RHEED patterns. There are two different kinds of clusters adsorbed on the surface. One is the bean-like cluster adsorbed on top of the \textquotedblleft 4$\times$\textquotedblright stripe, arranged in a perfect (6$\times$8) order. The other is the round cluster adsorbed between the \textquotedblleft 4$\times$\textquotedblright and \textquotedblleft 2$\times$\textquotedblright stripes with a lower density. It should be noted that in the high-resolution STM image, there are only 7 protrusions in one \textquotedblleft $\times$8\textquotedblright unit cell along [1$\overline{1}$0] either on the 2$\times$ or 4$\times$ stripes. The interspacing between two protrusions is uniformly $\sim$0.32~nm, showing the 8/7 \textquotedblleft magic\textquotedblright lattice match to the bulk truncated O atomic layer.

\begin {figure}[t]
 \includegraphics [width=3.0in,clip] {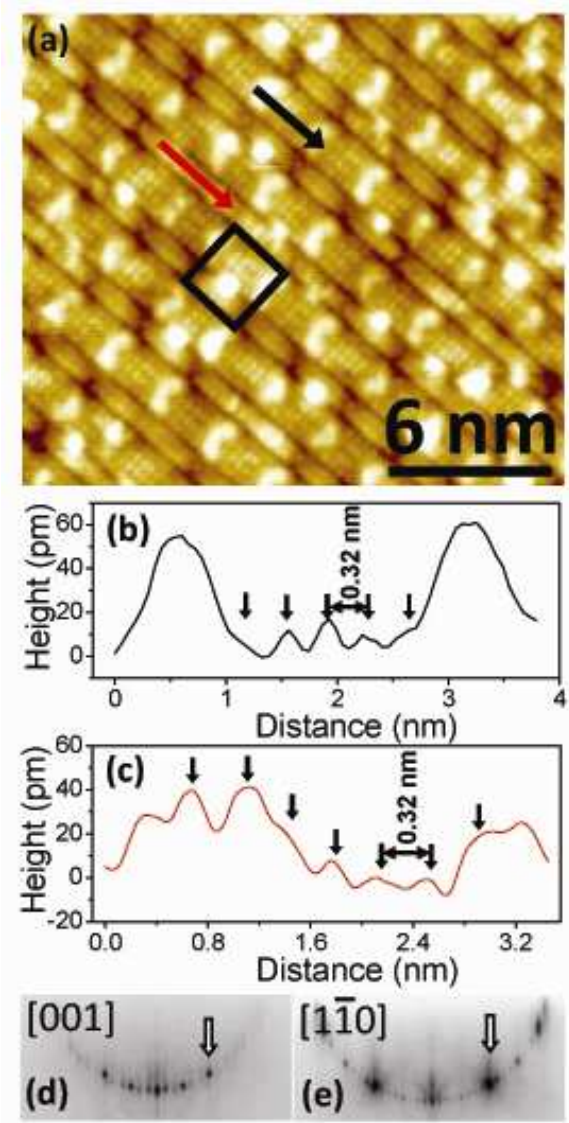}
 \caption{
(Color online) An STM image (1.0~V/20~pA) of the monophased (6$\times$8) surface. The unit cell is labeled with the rectangle. (b) and (c) Line profiles along the black and red (grey) arrows in (a), respectively. (d) and (e) The (6$\times$8) RHEED patterns along [001] and [1$\overline{1}$0] directions, respectively, with the integral diffractions indicated with the arrows.
}
\end{figure}

Figure 8 shows the surface after evaporating $\sim$0.05~ML Sr onto the monophased (6$\times$8) followed by annealing. Some \textquotedblleft 2$\times$\textquotedblright stripes condensate together forming the (2$\times$8) domains. The reversible structural phase transition between (2$\times$8) and (6$\times$8) can be realized by evaporating $\sim$0.1~ML Ti or Sr metal. 

\begin {figure}[t]
 \includegraphics [width=3.0in,clip] {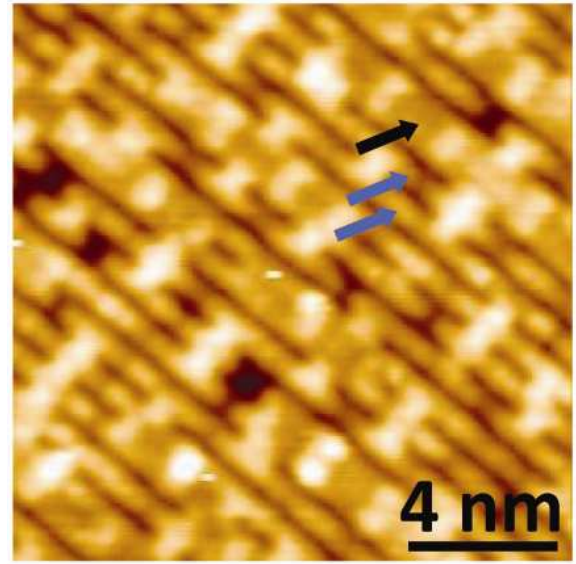}
 \caption{
(Color online) An STM image (1.5~V/20~pA) of the surface after evaporating $\sim$0.05~ML Sr onto the (6$\times$8) followed by annealing. The \textquotedblleft 4$\times$\textquotedblright and \textquotedblleft 2$\times$\textquotedblright stripes are indicated with black and blue (grey) arrows, respectively.
}
\end{figure}

\section{\label{sec:level2}Discussion} 

We establish a clear surface phase diagram in a broad range of chemical concentrations by evaporating Sr or Ti metals onto the SrTiO$_3$(110) surface followed by annealing. As the Ti dosage increases, the surface transforms from the (5$\times$1)-reconstructed SrTiO termination to the (4$\times$1), then to the (2$\times$8)-reconstructed O termination, and finally to the (6$\times$8). Reversible phase transitions can be induced by increasing Sr dosage. All the surface reconstructions are stable at high temperatures up to 1100 $^{\circ}$C and under oxygen partial pressure from UHV to 1$\times$10$^{-4}$ mbar. On the surface with mixed phases, the relative ratio of the two types of domains keeps constant as well within the ranges of temperature and oxygen partial pressure, without any other phase developed. Therefore it is concluded that thermodynamic equilibrium has been reached on the surface and the stabilization of the observed reconstructions is determined by the relative metal cation concentration, \textit{i.e.}, [Ti]/[Sr]. 

The surface composition can also be varied reversibly by different dose of Ar$^+$ sputtering followed by annealing, resulting in different areal ratio of SrTiO and O terminations, \textit{i.e.}, S$_A$ (see Fig.~2 and 3). This is also a thermodynamic process determined by [Ti]/[Sr] on the surface. As shown in Fig.~9, both S$_A$ and [Ti]/[Sr] on the annealed surface get saturated quickly after a few repeated treatment cycles with fixed parameters, unrelated to the initial configuration on the surface. Detailed high-resolution STM investigations actually reveal the reversible phase transitions between (5$\times$1), (4$\times$1), (2$\times$8), and (6$\times$8) by elaborately tuning the sputtering dose.

\begin {figure}[t]
 \includegraphics [width=3.4in,clip] {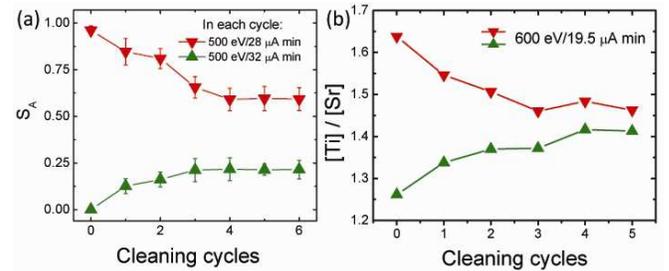}
 \caption{
(Color online) (a) S$_A$ of the SrTiO$_3$(110) surface with repeated cycles of cleaning (500 eV-sputtering followed by annealing). The statistics are done over 8 STM images (500$\times$500~nm$^2$) for each data point. (b) Surface concentration ratio [Ti]/[Sr] with different repeated cycles. The sputtering dose is fixed for each series of cleaning cycles, respectively. 
}
\end{figure}

\begin {figure}[b]
 \includegraphics [width=3.4in,clip] {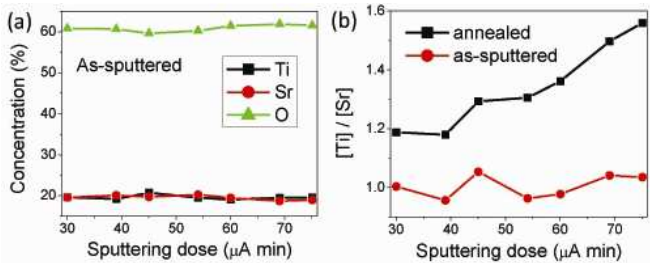}
 \caption{
(Color online) (a) The surface chemical concentration of the as-sputtered SrTiO$_3$(110) surface determined by XPS. (b) [Ti]/[Sr] ratio of the as-sputtered and annealed surface, respectively. Each data point in (a) and (b) is the saturated value after repeated cycles of treatment with the fixed sputtering dose, respectively.
}
\end{figure}

\begin {figure}[t]
 \includegraphics [width=3.4in,clip] {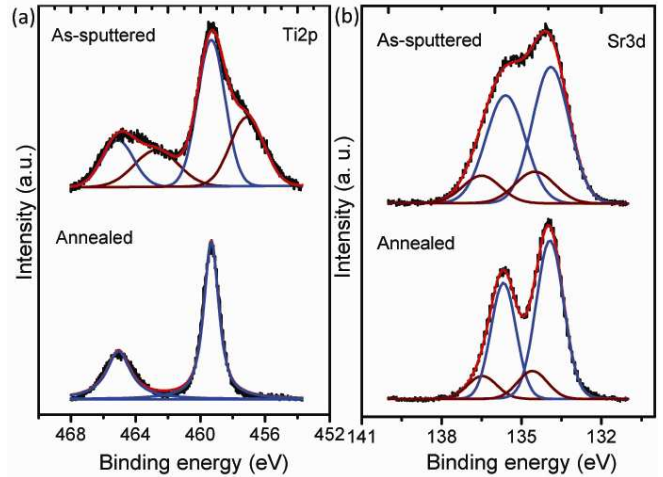}
 \caption{
(Color online) (a) The Ti~\textit{2p} core-level spectra of the annealed (lower panel) and as-sputtered (upper panel) surfaces, respectively. The spectrum taken on the annealed surface is fitted with a single doublet feature, while the one taken on the as-sputtered surface is fitted with an additional doublet feature shifted to lower binding energy by $\sim$1.65~eV. (b) The Sr~\textit{3d} core-level spectra of the annealed (lower panel) and as-sputtered (upper panel) surfaces, respectively.
}
\end{figure}

It has been proposed that the oxide polar surfaces can be stabilized by the nonstoichiometry \cite{RPP08 Noguera, PRB03 Noguera}. The chemical composition of the SrTiO$_3$ surface is related to various factors including thermal treatment, sputtering, adsorption, as well as crystalline orientation. For example, it is commonly agreed that Sr will be removed preferentially from the SrTiO$_3$(001) by sputtering, while on the (111)-orientated sample Ti will be preferentially removed \cite{JPCC08 Castell, Somorjai PRB78}. In contrast, on the SrTiO$_3$(110) surface in our experiments, we found that the atomic concentration ratio [Ti]/[Sr] is equal to 1 on the as-sputtered surface, and almost keeps constant for different sputtering dose, as shown in Fig.~10~(a). No preferential sputtering by Ar$^+$ has been observed within the ion energy range of 0.5 to 2~keV. However, the surface [Ti]/[Sr] ratio increases dramatically after annealing in relative to the as-sputtered sample, and the increment becomes larger monotonically as the sputtering dose increasing, as plotted in Fig.~10~(b). It is suggested that Ti (or Ti-O species) tend to diffuse towards the surface upon annealing to establish the thermodynamic equilibrium. Similarly, Ti-enriched nanostructures or even anatase TiO$_2$ islands have been obtained on the SrTiO$_3$(001) surfaces by intensive Ar$^+$ sputtering followed by annealing \cite{JACS03 Erdman, Nature02 Erdman, JPCC08 Castell, PRB07 Castell}. 

\begin {figure}[b]
 \includegraphics [width=3.4in,clip] {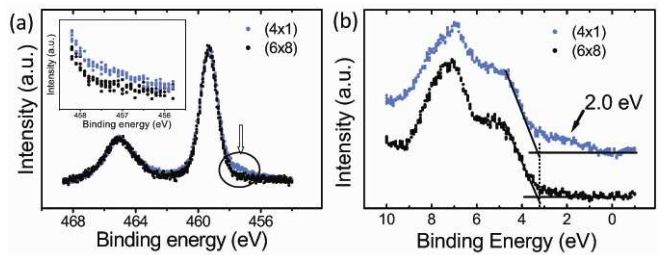}
 \caption{
(Color online) (a) The Ti~\textit{2p} core-level spectra taken on the monophased (4$\times$1) and (6$\times$8) surfaces, respectively. The inset zooms in their difference showing the appearance of the Ti$^{3+}$ component on the (6$\times$8) surface. (b) The valence-band spectra of the two reconstructions, respectively. The in-gap states are labeled with the arrow.
}
\end{figure}

In order to get the detailed information of such a picture, we carry out high-resolution XPS analyses on the (110) surface as presented in Fig.~11. Well-defined Ti~\textit{2p} and Sr~\textit{3d} core-level spectra are observed on the annealed surface. The Ti~\textit{2p} spectrum shows the single spin-orbital doublet feature of the Ti$^{4+}$ state as in the bulk crystal of SrTiO$_3$. The Sr~\textit{3d} spectrum can be fitted with two spin-orbital doublet pairs that originate from the different chemical coordination of Sr. The main doublet feature centered at $\sim$133~eV is the characteristic of the perovskite crystal, while the other feature shifting towards the higher energy by $\sim$0.9 eV can be attributed to Sr-O bond on the surface \cite{Szot, Yang, Courths}. Depending on different surface reconstructions, we detect a small change in the fine structure of Ti~\textit{2p} spectrum only. But dramatic difference is observed in the core-level spectra of both elements in comparison to the as-sputtered surface. Without annealing, the two peaks corresponding to the Ti~\textit{2p} single doublet state show broad shoulders at $\sim$~456.7 eV and 462.5 eV, respectively, indicating the appearance of Ti$^{3+}$ on the as-sputtered surface \cite{Webb}. This is consistent with the recent first-principles calculations that have revealed the existence of significant Ti antilike defects \cite{PRL09 M. Choi}. In the Sr~\textit{3d} spectrum, the weight of the high-energy doublet feature is higher than that on the annealed surface, consistently showing that the as-sputtered surface is deeply separated into SrO and TiO$_x$ chemical coordinations. 

\begin {figure}[t]
 \includegraphics [width=3.2in,clip] {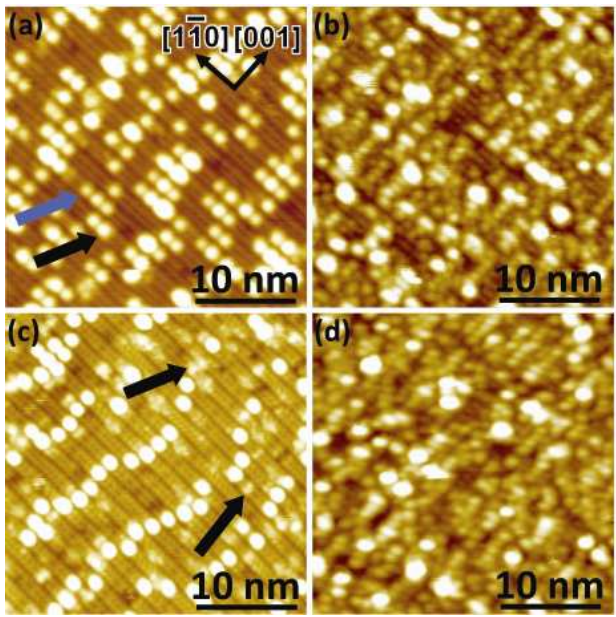}
 \caption{
(Color online) (a) and (b) STM images (1.0~V/20~pA) of the (4$\times$1) and (2$\times$8) surfaces evaporated with 0.02~ML Sr at room temperature, respectively. The original clusters and the evaporated Sr are indicated with black and blue (grey) arrows, respectively. (c) and (d) STM images (1.5~V/20~pA) of the (4$\times$1) and (2$\times$8) surfaces evaporated with 0.02~ML Ti at RT, respectively. The as-deposited Ti are indicated with black arrows.
}
\end{figure}

Considering the fact that titanium oxide have lower surface free energy as compared to the SrTiO$_3$ \cite{Nature02 Erdman, JACS03 Erdman, nm10 L.D. Marks}, the under-coordinated Ti induced by sputtering is driven to diffuse towards the surface upon annealing. Such a process is further assisted by the attractive interaction with the oxygen vacancies in the surface region that could also be induced by Ar$^+$ bombardment \cite{NM05 Kan}. Therefore the established thermodynamic equilibrium can be described as a Ti-enriched surface covering the stoichiometric SrTiO$_3$ bulk crystal. As the sputtering dose increases, the depth of the affected region increases and consequently the overall amount of Ti defects available to diffuse to the surface increases. Such an increase cannot be detected by XPS since the depth of XPS signal is much smaller than that of the region affected by sputtering. However, after annealing, the increment of the surface Ti concentration will be fully manifested by XPS within few top layers. On the other hand, it is commonly agreed that the formation energy and the migration barriers of oxygen vacancies are low in the bulk SrTiO$_3$ crystal. Therefore the oxygen ions diffuse to the surface upon annealing, keeping the metal cations sufficiently coordinated to reduce the surface energy. More importantly, such a mechanism determines the intrinsic dependence of the surface chemical composition of SrTiO$_3$(110) on the sputtering dose, offering the tunability of different reconstructions by adjusting the sputtering dose that is equivalent to evaporating Sr or Ti metal.

To further examine the properties of the SrTiO$_3$(110) surface with different reconstructions, the XPS core-level and valence-band spectra are carefully compared. Figure~12~(a) shows the Ti \textit{2p} core-level spectra taken from the monophased (4$\times$1) and (6$\times$8) surface, respectively. The (4$\times$1) surface shows mainly the Ti$^{4+}$ characteristic, while the (6$\times$8) surface also shows the existence of the Ti$^{3+}$ species as revealed by the low-energy shoulder \cite{Copel, Dawber, Henrich}. This is consistent with the chemical composition analyses that indicates the Ti enrichment on the O termination. Figure~12~(b) compares the valence-band spectra of the (4$\times$1)- and (6$\times$8)-reconstructed surface. The highest occupied state is determined by linearly extrapolating the onset edge of the signals in the spectra. The corresponding binding energy in relative to the Fermi energy (\textit{E}$_F$) is \textit{E}$_F$-\textit{E}$_{VB}$=3.3~eV. As the bandgap of the SrTiO$_3$ is \textit{E}$_g$=3.3~eV \cite{Cardona}, this value indicates that the \textit{E}$_F$ is near the minimum of the conduction band. In-gap states arise at $\sim$2.0~eV below \textit{E}$_F$ in the (6$\times$8) phase, indicating the existence of the Ti$^{3+}$ species \cite{Henrich, Nolan} that is consistent with the Ti~\textit{2p} core-level spectra. The appearance of the in-gap states also suggests that the O-terminated SrTiO$_3$(110) surface can be stabilized by filling electrons to the new states, in addition to the stoichiometry modification and adsorption of (charged) clusters as on the (4$\times$1) surface that also could be responsible for the charge compensation.

Since the surface electronic structure is varied with different reconstructions, the initial epitaxial growth behavior is also influenced. Figure~13 shows the different adsorption behaviors of Sr and Ti metals on the (4$\times$1)- and (2$\times$8)-reconstructed surfaces. With $\sim$0.02~ML Sr adsorbed onto the (4$\times$1) phase at RT, there are round protrusions appearing in the STM image, darker and smaller than those original clusters, as shown in Fig.~13~(a). By the statistics of the density of the adsorbed clusters, we found that Sr adsorbs as isolated single adatoms on top of the \textquotedblleft 4$\times$\textquotedblright stripes. Increasing the Sr dosage results in a higher density of the adatoms whose distribution still strictly follows the stripes. The Sr atoms adsorbed on the (2$\times$8) phase aggregate into randomly distributed clusters with nonuniform size. This might be the result of the strong interaction of Sr atoms with the (2$\times$8)-reconstructed surface due to its high reactivity. However, the adsorption behavior of Ti atoms is different from that of Sr adatoms. They aggregate into randomly distributed clusters with nonuniform size on the both (4$\times$1) and (2$\times$8) reconstructions.

\section{\label{sec:level2}Summary}

We establish a clear phase diagram of the SrTiO$_3$(110) surface, as shown in Fig.~14. The surface reconstruction phase can be tuned from the (5$\times$1)- to (4$\times$1)-reconstructed surface of the SrTiO termination, then to the (2$\times$8)- and finally to the (6$\times$8)-reconstructed surface of the O termination. All the structural phase transitions are reversibly driven by the change of the [Ti]/[Sr], which can be tuned by adjusting the sputtering dose or the amount of Sr/Ti adsorption. These two tuning methods are equivalent because the under-coordinated Ti induced by sputtering tend to diffuse towards the surface upon annealing. The electronic states of pure Ti$^{4+}$ valence are detected on the (4$\times$1) surface. In contrast, they are mixed with Ti$^{3+}$ on the (6$\times$8) surface. This causes the O termination more active than the (4$\times$1)-reconstructed SrTiO termination, resulting in the different absorption behaviors during the initial epitaxial growth.

\begin {figure}[t]
 \includegraphics [width=3.0in,clip] {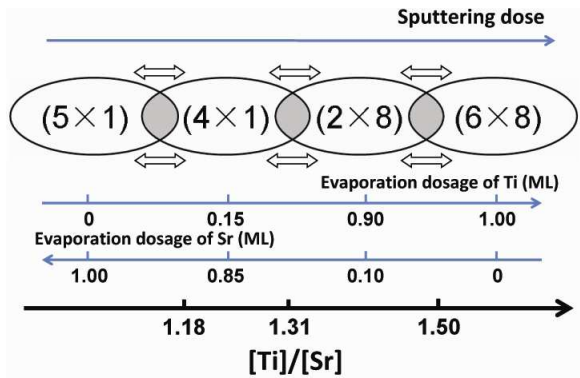}
 \caption{
(Color online) The phase diagram of the SrTiO$_3$(110) surface in relative to the surface [Ti]/[Sr] ratio that can be tuned by the sputtering dose or the evaporation of Sr/Ti.
}
\end{figure}

\begin{acknowledgments}
This work is supported by Chinese NSF-10704084 and Chinese MOST (2006CB921300 \& 2007CB936800).
\end{acknowledgments}


\begin{thebibliography}{}
\bibitem{sensitivity factors} The sensitivity factors are $\Sigma_{Sr}$=1.843, $\Sigma_{Ti}$=2.001, and $\Sigma_O$=0.711, respectively.
\end{thebibliography}
\end{document}